\documentclass[letterpaper,11pt]{article}

\pdfoutput=1
\usepackage{jheppub}
\pdfoutput=1
\usepackage{amsmath}
\usepackage{graphicx}
\usepackage{xspace}
\usepackage{slashed}
\usepackage{xcolor}
\usepackage[normalem]{ulem}

\newcommand{\red}[1]{{\color{red}{#1}}}

\usepackage{jheppub}
\usepackage{multirow}

\usepackage{subfig}
\usepackage{xspace}
\usepackage[countmax]{subfloat}

\usepackage{amssymb}
\usepackage{amsmath}
\usepackage{cancel}
\usepackage{tabu,booktabs}
\usepackage{color}
\usepackage{braket}
\usepackage{graphicx}
\usepackage{multirow}
\usepackage{verbatim}
\usepackage{amsthm}
\usepackage{slashed}
\usepackage{wasysym}
\usepackage{simplewick}
\usepackage{mathtools}
\usepackage{soul}
\usepackage{xspace}

\definecolor{dancomment}{RGB}{0,159,0}

\def\cL{\mathcal{L}}
\def\cM{\mathcal{M}}

\def\cO{\mathcal{O}}

\def\tr{{\rm tr}}

\newcommand{\Tau}{\mathcal{T}}

\def\nn{{\nonumber}}

\newcommand{\Eq}[1]{Equation~\eqref{#1}}

\DeclareRobustCommand{\Sec}[1]{Sec.~\ref{#1}}

\DeclareRobustCommand{\Tab}[1]{Table~\ref{#1}}

\DeclareRobustCommand{\Fig}[1]{Fig.~\ref{#1}}

\DeclareRobustCommand{\Eq}[1]{Eq.~(\ref{#1})}
\DeclareRobustCommand{\Eqs}[2]{Eqs.~(\ref{#1}) and (\ref{#2})}

\DeclareRobustCommand{\Ref}[1]{Ref.~\cite{#1}}

\def\be{\begin{equation}}
\def\ee{\end{equation}}

\newcommand{\SCETi}{\mbox{${\rm SCET}_{\rm I}$}\xspace}
\newcommand{\SCETii}{\mbox{${\rm SCET}_{\rm II}$}\xspace}

\allowdisplaybreaks[3]

\newcommand{\eq}[1]{Eq.~\eqref{eq:#1}}
\newcommand{\eqs}[2]{Eqs.~\eqref{eq:#1} and \eqref{eq:#2}}

\renewcommand{\sec}[1]{Sec.~\ref{sec:#1}}

\newcommand{\abs}[1]{\lvert#1\rvert}
\newcommand{\ord}[1]{\mathcal{O}(#1)}

\newcommand{\df}{\mathrm{d}}

\newcommand\bn{{\bar n}}

\newcommand{\eps}{\epsilon}

\newcommand{\fd}[2]{\parbox{#1}{\includegraphics[width=#1]{#2}}}

\newcommand{\hH}{\widehat{H}}
\newcommand{\hS}{\widehat{S}}

\newcommand{\Ecm}{E_\mathrm{cm}}

\newcommand{\BornM}{\hat{\cM}_\text{born}}

\newcommand{\pstar}{\tilde{P}}

\newcommand{\collin}[2]{{#1 \parallel #2}}
\newcommand{\hel}{h}
\newcommand{\Msquared}{(|A|^{2})}
\newcommand{\colllabel}{j}

  \newcount\hour \newcount\minute
  \hour=\time \divide \hour by 60 \minute=\time
  \newcommand{\todaytime}{\today \ -- \number\hour :\ifnum \minute<10 0\fi\number\minute}

\preprint{MIT-CTP 5090}

\title{Helicity Methods for High Multiplicity Subleading Soft and Collinear Limits}

\author[1]{Arindam Bhattacharya,}
\author[2,3]{Ian Moult,}
\author[1]{Iain W. Stewart,}
\author[1]{and Gherardo Vita}

\affiliation[1]{Center for Theoretical Physics, Massachusetts Institute of Technology, Cambridge, MA 02139, USA}
\affiliation[2]{Berkeley Center for Theoretical Physics, University of California, Berkeley, CA 94720, USA}
\affiliation[3]{Theoretical Physics Group, Lawrence Berkeley National Laboratory, Berkeley, CA 94720, USA}

\emailAdd{arindamb@mit.edu}
\emailAdd{ianmoult@lbl.gov}
\emailAdd{iains@mit.edu}
\emailAdd{vita@mit.edu}

\abstract{The factorization of multi-leg gauge theory amplitudes in the soft and collinear limits provides strong constraints on the structure of amplitudes, and enables efficient calculations of multi-jet observables at the LHC. There is significant interest in extending this understanding to include subleading powers in the soft and collinear limits. 
While this has been achieved for low point amplitudes, for higher point functions there is a proliferation of variables and more complicated phase space, making the analysis more challenging.
By combining the subleading power expansion of spinor-helicity variables in collinear limits with consistency relations derived from the soft collinear effective theory, we show how to efficiently extract the subleading power leading logarithms of $N$-jet event shape observables directly from known spinor-helicity amplitudes. At subleading power, we observe the presence of power law singularities arising solely from the expansion of the amplitudes, which for hadron collider event shapes lead to the presence of derivatives of parton distributions.  The techniques introduced here can be used to efficiently compute the power corrections for $N$-jettiness subtractions for processes involving jets at the LHC.
}

\begin{document} 

\maketitle

\section{Introduction}\label{sec:intro}

The factorization properties of multi-leg gauge theory amplitudes in the soft and collinear limits are essential for our theoretical understanding of these amplitudes, as well as for the calculation of multi-jet observables at hadron colliders. While the leading power soft and collinear limits have been extensively studied, very little is known about the subleading power factorization properties of multi-leg amplitudes, or multi-jet observables.

There has recently been significant progress in understanding the structure of power corrections in the soft and collinear limits \cite{Manohar:2002fd, Beneke:2002ph, Pirjol:2002km, Beneke:2002ni, Bauer:2003mga, Hill:2004if,
Mannel:2004as, Lee:2004ja, Bosch:2004cb, Beneke:2004in, Tackmann:2005ub,
Trott:2005vw,Dokshitzer:2005bf,Laenen:2008ux,Laenen:2008gt,Paz:2009ut,Benzke:2010js,Laenen:2010uz,Freedman:2013vya,Freedman:2014uta,Bonocore:2014wua,Larkoski:2014bxa,Bonocore:2015esa,Kolodrubetz:2016uim,Bonocore:2016awd,Moult:2016fqy,Boughezal:2016zws,DelDuca:2017twk,Balitsky:2017flc,Moult:2017jsg,Goerke:2017lei,Balitsky:2017gis,Beneke:2017ztn,Feige:2017zci,Moult:2017rpl,Chang:2017atu,Boughezal:2018mvf,Ebert:2018lzn,Bahjat-Abbas:2018hpv}, including the first all order resummation of power suppressed logarithms for collider observables with soft and collinear radiation \cite{Moult:2018jjd} and more recently for the case of threshold \cite{Beneke:2018gvs}. However, complete calculations of the all orders structure of power suppressed terms have so far focused on the case of two back-to-back jets, corresponding to color singlet production at the LHC, or dijet production in $e^+e^-$. Both for improving our theoretical understanding, as well as for practical applications for observables at the LHC, it is important to be able to extend these calculations to the multi-jet case.

Compact expressions for multi-point amplitudes are typically expressed using the spinor-helicity formalism \cite{DeCausmaecker:1981bg,Berends:1981uq,Gunion:1985vca,Xu:1986xb}, and color ordering techniques \cite{Berends:1987me,Mangano:1987xk,Mangano:1988kk,Bern:1990ux}. See e.g. \cite{Dixon:1996wi,Dixon:2013uaa} for reviews. Due to the success of unitarity  \cite{Bern:1994zx,Bern:1994cg}  and recursion \cite{Britto:2004ap,Britto:2005fq} based techniques, a wealth of tree, one- and two-loop multi-point amplitudes are known in QCD. However, for the most part, this wealth of data has not been exploited in the study of subleading power corrections to collider observables.

In this paper we provide a method to directly and efficiently compute subleading power logarithms for multi-jet event shape observables using known spinor amplitudes. First, we study the expansion of the two-particle collinear limit to subleading powers in terms of spinor helicity variables, providing a convenient parametrization in terms of standard kinematic variables. Then, we exploit consistency relations derived in soft collinear effective theory (SCET) \cite{Bauer:2000ew, Bauer:2000yr, Bauer:2001ct, Bauer:2001yt, Bauer:2002nz} to show that the leading logarithms at subleading power for a broad class of multi-jet event shape observables can be computed using only the two-particle collinear limit, to any order in $\alpha_s$. The two particle collinear limit is particularly convenient from the perspective of multi-jet calculations, since it avoids the complicated phase space integrals that appear in soft limits. We use several simple examples to show explicitly how this can be done in an efficient manner. These techniques should enable a rapid extension of the availability of power corrections to multi-jet processes.

By extending to the multi-point case, we are also able to improve our theoretical understanding of subleading power corrections and factorization, since features that are specific to two back-to-back jet directions no longer apply. In particular, we observe that at subleading powers, generic multipoint amplitudes exhibit power law, instead of logarithmic divergences. The proper treatment of these power law divergences in terms of distributions leads to derivatives of the parton distribution functions (PDFs) in hadron collider observables. An interesting feature about multi-point amplitudes is that these singularities arise already at the squared amplitude level, even if the corresponding phase space integrals are not themselves singular. This is a generic features whose treatment at fixed order provides the first step towards understanding their all orders structure for generic amplitudes. 

An outline of this paper is as follows. In \Sec{sec:spinor} we discuss the parametrization of the two-particle collinear limit in spinor-helicity variables, showing how we can efficiently expand amplitudes to subleading powers in the collinear limit, and giving several concrete examples. In \Sec{sec:log} we show how we can use consistency relations derived in SCET to extract subleading power logarithms for multi-jet event shape observables from the two-particle collinear limit. In \Sec{sec:powlaw} we discuss the treatment of power law divergences which appear in the power expansion of amplitudes. We conclude in \Sec{sec:conclusions}, and provide an outlook for a number of applications of the techniques discussed here.

\section{Subleading Power Expansions of Spinors}\label{sec:spinor}

In this section, we describe in detail the subleading power expansion of spinor helicity variables, focusing on the behavior and parametrization of the two particle collinear limit at subleading powers. While soft limits have been studied extensively (see e.g. \cite{Strominger:2017zoo} for a recent review), subleading power collinear limits are much less well studied, and therefore parametrizations of spinors in these limits are less widely known in the literature. A convenient parametrization of the two particle collinear limit at subleading powers was given in \cite{Stieberger:2015kia,Nandan:2016ohb}. In this section, we generalize this parametrization, and make it explicit in terms of the standard momenta that are useful for calculations of observables in the collinear limit. In \Sec{sec:log} we will apply this expansion to extract subleading power logarithms in event shape observables.

\subsection{Subleading Power Collinear Limit}\label{sec:toolbox}

Here we will consider the subleading power expansion of the two-particle collinear limit. We assume that we have two particles with momenta $p_1$ and $p_2$ that are collinear along a direction $n$. It is convenient to work in lightcone coordinates, decomposing a given momentum $k$ as $(n\cdot k, \bn \cdot k, k_\perp) \equiv (k^+, k^-, k_\perp)$. Here $\bar n$ is an auxiliary lightlike vector.  As a concrete example we can take the vectors to be $n^\mu =(1,0,0,1)$ and $\bn^\mu=(1,0,0,-1)$. We then define particles collinear to the $n$ direction to have the momentum scaling
\begin{align} \label{eq:collinear}
(k^-,k^+,k_{\perp})\sim Q(\lambda^0,\lambda^2,\lambda)\,,
\end{align}
where $Q$ is some typical hard scale for the energy of the collinear radiation and $\lambda \ll 1$ is our power expansion parameter. Note that $\lambda$ is a scaling parameter that determines the size of various contributions, and hence does not itself show up in expanded amplitudes.
With this momentum scaling, it is straightforward to expand amplitudes expressed in terms of standard Mandelstam invariants. However, we would also like to be able to expand amplitudes expressed in terms of spinor helicity variables. We follow the notation of~\cite{Dixon:1996wi}.

To expand particles $1$ and $2$ in the two particle collinear limit, we parametrize the full spinors as
\begin{align}\label{eq:def}
|1\rangle &=c\ |p\rangle - \epsilon s\ |r\rangle\,,\\
|2\rangle &=s\ |p\rangle + \epsilon' c\ |r\rangle\,,
 \nn
\end{align}
where $p$ and $r$ are momenta along $n$ and $\bar{n}$ respectively, and 
the $|p\rangle$ term dominates.  The parameters $c$ and $s$ are such that $c^2+s^2=1$,  and $\epsilon$ and $\epsilon'$ are complex parameters involving small combinations of momenta in which we will expand, with $\epsilon,\epsilon'\sim \lambda$. Both $\epsilon$ and $\epsilon'$ are needed in order to take generic collinear limits.  The special case with $\epsilon=\epsilon'$ corresponds to an additional kinematic restriction (discussed below), in which case \eq{def} is identical to the decomposition in 
Refs.~\cite{Stieberger:2015kia,Nandan:2016ohb}.
For square brackets we have the analogous decomposition
\begin{align} \label{eq:sqdef}
|1] &=c\ |p] - \epsilon^{*} s\ |r]\,,\\
|2] &=s\ |p] + \epsilon^{\prime *} c\ |r]\,,
 \nn
\end{align}
where again the $|p]$ term dominates, and the $^*$ indicates complex conjugation.
Inverting \Eqs{eq:def}{eq:sqdef}, one obtains
\begin{align}\label{eq:sys1}
|p\rangle &=\frac{\epsilon' c}{\epsilon' c^2 + \epsilon s^2}\: |1\rangle 
  + \frac{s\epsilon}{\epsilon' c^2 + \epsilon s^2} \: |2\rangle
  \,,
& |p] &=\frac{\epsilon^{\prime *} c}{\epsilon^{\prime *} c^2 + \epsilon^* s^2}\: |1] 
  + \frac{s\epsilon^*}{\epsilon^{\prime *} c^2 + \epsilon^* s^2} \: |2]
  \,,
 \\ 
|r\rangle &=  \frac{c}{\epsilon' c^2 + \epsilon s^2}\: |2\rangle 
    -\frac{s}{\epsilon' c^2 + \epsilon s^2}\: |1\rangle \,,
& |r] &=  \frac{c}{\epsilon^{\prime *} c^2 + \epsilon^* s^2}\: |2]
    -\frac{s}{\epsilon^{\prime *} c^2 + \epsilon^* s^2}\: |1] \,.
 \nn
\end{align}

Now, we solve for the quantities $p,r,c,s,\epsilon,\epsilon^{\prime}$ in terms of $p_1,p_2$.   Without loss of generality, 
we shall take $n$ and $\bar{n}$ to be the four vectors $(1,\hat{n})$ and $(1,-\hat{n})$ and use the following representations for the spinors \cite{Dixon:1996wi}
\begin{align}\label{eq:dirnot}
|k\rangle&=\frac{1}{\sqrt{2}}\begin{pmatrix}
\sqrt{k^-}\\
\sqrt{k^+}e^{i\phi_k}\\
\sqrt{k^-}\\
\sqrt{k^+}e^{i\phi_k}
\end{pmatrix}\,,&
|k]&=\frac{1}{\sqrt{2}}\begin{pmatrix}
\sqrt{k^+}e^{-i\phi_k}\\
-\sqrt{k^-}\\
-\sqrt{k^+}e^{-i\phi_k}\\
\sqrt{k^-}
\end{pmatrix}\,,
 & e^{i\phi_k}& =\frac{k^x+ik^y}{\sqrt{k^+ k^-}} \,.
\end{align}
These correspond to using the Dirac basis of the gamma matrices, and by default we assume that the convention for spinor momentum labeling is always outgoing. Thus these momenta are positive for outgoing particles and negative for incoming particles.\footnote{For  incoming particles $\sqrt{-k^\pm}=i\sqrt{k^\pm}$, and conjugated spinors also get an extra minus sign, which ensures that spinor identities are valid for both positive and negative outgoing momenta.} 
Solving \eq{sys1} to obtain the $p,r,c,s,\epsilon,\epsilon^\prime$ we obtain
\begin{align}
p &= \left(p_1^-+p_2^-\right)\frac{n}{2}\,,&
r &=  \left(p_1^-+p_2^-\right)\frac{\bar{n}}{2}\,,
 \\
c &= \sqrt{\frac{p_1^-}{p_1^-+p_2^-}}\equiv\sqrt{x}\,,&
s &= \sqrt{\frac{p_2^-}{p_1^-+p_2^-}}=\sqrt{1-x}\,,
 \nn \\
\epsilon &= -\sqrt{\frac{p_1^+}{p_2^-}}\ e^{i\phi_1} 
  \equiv -\zeta\, e^{i\phi_1} \,,&
\epsilon^\prime &= \sqrt{\frac{p_2^+}{p_1^-}}\ e^{i\phi_2}
  \equiv \zeta'\, e^{i\phi_2}  \,,
  \nn
\end{align}
where $e^{i\phi_j}=(p_j^x+ip_j^y)/\sqrt{p_j^+p_j^-}$ for $j=1,2$, and $\epsilon\sim\epsilon'\sim \lambda$ are the expansion parameters.

Note that the scaling of collinear momenta makes it manifest that $p$, $r$, $c$, and $s$ are $\mathcal{O}(\lambda^0)$ quantities, and thus \Eq{eq:def}, allows us to safely expand amplitudes as a power series in $\epsilon$ and $\epsilon'$ which are the only ${\cal O}(\lambda)$ variables. One also observes the appearance of the energy fractions $x$ and $(1-x)$ of 1 and 2 respectively. 
The series thus obtained will be the limit of the amplitude when particles $1,2$ become collinear.

For only final state collinear particles (or only initial state collinear particles), we can exploit the freedom of choosing $\hat n$ in order to make the total transverse momentum for all particles that are being taken in the collinear limit of \eq{collinear}, to be zero. For two final state particles this implies $p_1^\perp = -p_2^\perp$. Here $p_1^->0$, $p_2^->0$, and we define the momentum fraction as
\begin{align}
x\equiv \frac{p_1^-}{p_1^-+p_2^-} \,,
\end{align}
where $0\le x\le 1$. 
With these assumptions, 
\begin{align}
 c&=\sqrt{x} \,, &s& = \sqrt{1-x} \,,
 \\
e^{i\phi}&=e^{i\phi_2}=-e^{i\phi_1}  \,,
 & \zeta &= \zeta' \,, 
 &\epsilon &= \zeta\, e^{i\phi} = \epsilon^{\prime}
\,.
  \nn
\end{align}
In this case the exact spinor decompositions become simpler:
\begin{align}\label{eq:deffinal}
|1\rangle &=c\ |p\rangle - \epsilon s\ |r\rangle\,,
 & |p\rangle &=c\ |1\rangle + s\ |2\rangle\,,
 \\
|2\rangle &=s\ |p\rangle + \epsilon c\ |r\rangle\,,
 & \epsilon\: |r\rangle &=-s\ |1\rangle + c\ |2\rangle
 \nn \,,
 \\
|1] &=c\ |p] - \epsilon^{*} s\ |r]\,,
 & |p] &=c\ |1] + s\ |2]\,,
 \nn \\
|2] &=s\ |p] + \epsilon^{*} c\ |r]\,,
 & \epsilon^{*}\, |r] &=-s\ |1] +  c\ |2] 
 \,. \nn
\end{align}
Since here $\epsilon'=\epsilon$, the expansion has also been reduced to a single small parameter $\epsilon$.

Another interesting case is the collinear limit between an emission with outgoing momentum $p_1$ and an initial state particle with outgoing momentum $p_2$. Here $p_1^->0$ and $p_2^-<0$ and we can define the momentum fraction $(1-z)$ for the emission relative to the initial particle, via
\begin{align}
 1 - z = \frac{p_1^-}{-p_2^-} \,,
\end{align}
where $0\le z \le 1$. 
In this case its natural to choose $\hat n$ so that we have $p_2^\perp=0$ (rather than the sum of the two $\perp$-momenta).  With these assumptions we have
\begin{align}
  c &= \sqrt{1-\frac{1}{z}} = i \sqrt{\frac{1}{z} -1} \,,
 &s & = \frac{1}{\sqrt{z}} \,,
 &  \epsilon' & = \zeta' e^{i\phi_2} = 0 \,.
\end{align}
We also have $|2\rangle = (1/\sqrt{z}) |p\rangle$, and $|2]=(1/\sqrt{z}) |p]$, so that these spinors are already aligned with the collinear direction. In this situation there is still an expansion for $|1\rangle$ and $|1]$ from \eqs{def}{sqdef}, and once again, the expansion is in the single parameter $\epsilon$.

Employing these parametrizations of the spinors, we can efficiently expand in the two particle collinear limits.
The usual leading power simplification that arises for an amplitude in the collinear limit can be illustrated with the MHV four-point gluon amplitude. In the limit where $12$ are collinear we have
\begin{align} \label{eq:MHVexpn}
 A(1^-2^+3^-4^+)_{1\parallel 2} 
 =  \frac{\langle 13\rangle^4}
  {\langle 12\rangle \langle 23 \rangle \langle 34 \rangle \langle 41 \rangle} 
  \Big|_{1\parallel 2}
  &= \frac{c^3}{s\langle 12\rangle} \frac{ \langle p3\rangle^4}
  { \langle p3 \rangle \langle 34 \rangle \langle 4p \rangle} 
   + \ldots \,,
\end{align}
where the splitting function $c^3/(s\langle 12) \rangle) =c^3/(s\epsilon \langle pr\rangle) \sim \lambda^{-1}$ makes the displayed term ${\cal O}(\lambda^{-1})$. This result is valid for both outgoing and incoming particles. The terms in the ellipses in \eq{MHVexpn} are terms of higher power in the collinear limit.
In the next few sections we illustrate results for subleading terms in the collinear limit through a couple of examples.

\subsection{Example: $H\to \bar qq \bar Q Q$}

As an illustrative example, we shall derive the subleading collinear limits for the process of decay of a color singlet into 4 partons, which has all particles outgoing. For concreteness and simplicity, we shall take the singlet to be a Higgs, and the 4 partons to be two quark-antiquark pairs with differing flavors. At tree-level, only the following helicity confugurations contribute \cite{Kauffman:1996ix}:
\begin{align}\label{eq:hqqQQamp}
 A(1_q^+,2_{\bar{q}}^-;3_Q^+,4_{\bar{Q}}^-;5_H)=\frac{1}{2}\left(\frac{\langle 24\rangle^2}{\langle 12\rangle\langle 34\rangle}+\frac{[13]^2}{[12][34]}\right)\,,\\
 A(1_q^+,2_{\bar{q}}^-;3_Q^-,4_{\bar{Q}}^+;5_H)=-\frac{1}{2}\left(\frac{\langle 23\rangle^2}{\langle 12\rangle\langle 34\rangle}+\frac{[14]^2}{[12][34]}\right)\,.
  \nn
\end{align}
 The conjugate helicity configurations can be obtained using parity. To illustrate the types of structures that the subleading power expansion yields we will consider two choices for the pairs of particles going collinear. In one case there will be no leading power collinear limit, and in the other case there is a leading power collinear, which gives a more complicated result.
 
 We begin by analyzing the behavior of the amplitude when quark 1 and antiquark 4 become collinear. This particular collinear limit has no leading power, $\mathcal{O}(\lambda^{-1})$, term in the amplitude since there is no spinor product with $14$ in the denominator of \eq{hqqQQamp}. This makes extracting the next-to-leading behavior in the collinear limit straightforward, since one can just use the standard leading power expressions for the spinors, namely
 \begin{align}
|1\rangle &=\sqrt{x}\ |p\rangle\,,&
|4\rangle &=\sqrt{1-x}\ |p\rangle\,,\\
|1] &=\sqrt{x}\ |p]\,,&
|4] &=\sqrt{1-x}\ |p]\,. \nn
\end{align} 
 Substituting these into the amplitudes, we get the following expansion in $\lambda$
 \begin{align}
 A(1_q^+,2_{\bar{q}}^-;3_Q^+,4_{\bar{Q}}^-;5_H)_{\collin{1}{4}}&=0\times\mathcal{O}(\lambda^{-1})-\frac{1}{2}\left(\sqrt{\frac{x}{1-x}}\ \frac{\langle p2\rangle}{\langle p3\rangle}+\sqrt{\frac{1-x}{x}}\ \frac{[3p]}{[2p]}
 \right)+\mathcal{O}(\lambda)\,,
  \nn \\
 A(1_q^+,2_{\bar{q}}^-;3_Q^-,4_{\bar{Q}}^+;5_H)_{\collin{1}{4}}&=0\times\mathcal{O}(\lambda^{-1})+\frac{\langle 23\rangle^2}{2\sqrt{x(1-x)}\langle p2\rangle\langle p3\rangle}+\mathcal{O}(\lambda)\,.
 \end{align}
Note that these results are expressed entirely in terms of the collinear spinors $|p\rangle$, $|p]$, the momentum fraction $x$, and the spinors for the other directions.  
These subleading power expressions take a very simple form, due to the fact that there was no leading power term. In the first case where the quark and antiquark have opposite helicities, we see that the amplitude behaves like that for a scalar in the direction $p$. In the second case when they have the same helicity, it behaves like an amplitude for a particle with spin 1 along $p$. It would be interesting to understand this in more generality. Some work in this direction, involves representing subleading power collinear limits of gluon amplitudes in terms of mixed Einstein-Yang-Mills amplitudes \cite{Stieberger:2015kia}.
 
For helicity configurations that do not have a leading power limit, it is also simple to get the power suppressed squared amplitude in the collinear limit to $\mathcal{O}(\lambda^2)$, since this comes only from the interference of the two $\mathcal{O}(\lambda)$ suppressed amplitudes. Neglecting any color structures, we find that the amplitude squared have the following subleading $\mathcal{O}(\lambda^0)$ terms:
\begin{align}\label{eq:hqqQQsq}
 |A(1_q^+,2_{\bar{q}}^-;3_Q^+,4_{\bar{Q}}^-;5_H)|^2_{\collin{1}{4}}&=0\times\mathcal{O}(\lambda^{-2})+0\times\mathcal{O}(\lambda^{-1})+\frac{\left[(1-x)\ s_{p2}+x\ s_{p3}\right]^2}{4\ x(1-x)s_{p2}s_{p3}}+\mathcal{O}(\lambda)\,,
 \nn \\
 |A(1_q^+,2_{\bar{q}}^-;3_Q^-,4_{\bar{Q}}^+;5_H)|^2_{\collin{1}{4}}&=0\times\mathcal{O}(\lambda^{-2})+0\times\mathcal{O}(\lambda^{-1})+\frac{s_{23}^2}{4\ x(1-x)s_{p2}s_{p3}}+\mathcal{O}(\lambda)\,.
\end{align} 
In this case, they involve only Mandelstam invariants with the direction $p$, as well as the momentum fraction $x$, but do not otherwise involve the substructure of the splitting. 
These can now be trivially integrated over the collinear phase space to obtain subleading power corrections for an event shape observable, as we will describe in \Sec{sec:log}.
 
 The previous limit was particularly simple due to the fact that it did not have a leading power term. To illustrate a slightly more complicated example, we examine the behavior of the amplitudes in (\ref{eq:hqqQQamp}) when the $12$ quarks become collinear. This collinear limit has a leading power term, which is governed by the standard leading power collinear factorization. We must therefore keep all the subleading terms in the expansion of the spinors. In this case the required substitutions are
\begin{align}
	\langle 12\rangle&=\zeta e^{i\phi}\langle pr\rangle\,,&
	[12]&=\zeta e^{-i\phi}[pr]\,,
   \\
	\langle 1i\rangle&=\sqrt{x}\ \langle pi\rangle-\zeta e^{i\phi}\sqrt{1-x}\ \langle ri\rangle\,,&
	[1i]&=\sqrt{x}\ [pi]-\zeta e^{-i\phi}\sqrt{1-x}\ [ri],\ \text{for}\ i=3,4\,,
   \nn \\
	\langle 2i\rangle&=\sqrt{1-x}\ \langle pi\rangle+\zeta e^{i\phi}\sqrt{x}\ \langle ri\rangle\,,& 	
	[2i]&=\sqrt{1-x}\ [pi]+\zeta e^{-i\phi}\sqrt{x}\ [ri],\ \text{for}\ i=3,4
   \,. \nn
\end{align}
Plugging these in to the amplitudes and expanding, we arrive at the following structure for the amplitudes 
\begin{align}
	A=A^{(0)}+A^{(1)}+A^{(2)}+\cdots\,,
\end{align}
where the leading power term $A^{(0)}\sim\mathcal{O}(\lambda^{-1})$, and each successive term acquires a power in $\lambda$, so $A^{(n)}\sim\mathcal{O}(\lambda^{-1+n})$.

The leading power amplitudes obey the well known factorization into a splitting function and a lower point amplitude
\begin{align}\label{eq:LPsplitfact}
	A^{(0)}=\sum_{\hel=\pm}\text{Split}_{-\hel}(a,b;x) A_{n-1}(\ldots, p^\hel, \ldots)\,.
\end{align}
where the tree level splitting amplitudes can be found summarized in Appendix II of Ref.~\cite{Bern:1994zx}.
For our example, the lower point amplitudes we require are
\begin{align}
	A(p^+;3_q^+,4_{\bar q}^-;5_H)=\frac{[p3]^2}{[34]}\,, \qquad
	A(p^-;3_q^+,4_{\bar q}^-;5_H)=\frac{\langle p4\rangle^2}{\langle34\rangle}\,,
\end{align}
and the relevant splitting functions are given by 
\begin{align}
	\text{Split}_{+}(q,\bar{q};x)=\frac{(1-x)}{\langle q\bar{q}\rangle}\,, \qquad \text{Split}_{-}(q,\bar{q};x)=\frac{x}{[q\bar{q}]}\,.
\end{align}
We therefore have
\begin{align}
	A^{(0)}(1_q^+,2_{\bar{q}}^-;3_Q^+,4_{\bar{Q}}^-;5_H)_{\collin{1}{2}}&=\frac{(1-x)e^{-i\phi}\ \langle p4\rangle^2}{2 \zeta \langle pr\rangle\langle 34\rangle}+\frac{x\ e^{i\phi}[3p]^2}{2\zeta [rp][43]}\\
	&=\frac{(1-x)e^{-i\phi}}{2\zeta \langle pr\rangle}\ A(p^-;3_Q^+,4_{\bar Q}^-;5_H)+\frac{x\ e^{i\phi}}{2\zeta [pr]}\ A(p^+;3_Q^+,4_{\bar Q}^-;5_H)\,,\nn
\end{align}
and
\begin{align}
	A^{(0)}(1_q^+,2_{\bar{q}}^-;3_Q^-,4_{\bar{Q}}^+;5_H)_{\collin{1}{2}}&=-\bigg[\frac{(1-x)e^{-i\phi}\ \langle p3\rangle^2}{2 \zeta\langle pr\rangle\langle 34\rangle}+\frac{x\ e^{i\phi}[4p]^2}{2\zeta[rp][43]} \bigg] \\
	&= \frac{(1-x)e^{-i\phi}}{2\zeta\langle pr\rangle}\ A(p^-;3_Q^-,4_{\bar Q}^+;5_H)+\frac{x\ e^{i\phi}}{2\zeta[pr]}\ A(p^+;3^-_{Q},4^+_{\bar Q}) \nn \,,
\end{align}
which implies
\begin{align}
	A^{(0)}(1_q^+,2_{\bar{q}}^-;3_Q^+,4_{\bar{Q}}^-;5_H)_{\collin{1}{2}}&=\sum_{\hel=\pm}\text{Split}_{-\hel}(1_q^+,2_{\bar{q}}^-;x)\ A(p^{h};3_Q^+,4_{\bar Q}^-;5_H), \\
	A^{(0)}(1_q^+,2_{\bar{q}}^-;3_Q^-,4_{\bar{Q}}^+;5_H)_{\collin{1}{2}}&=\sum_{\hel=\pm}\text{Split}_{-\hel}(1_q^+,2_{\bar{q}}^-;x)\ A(p^{h};3_Q^-,4_{\bar Q}^+;5_H)\,,\nn
\end{align}
as expected from \eq{LPsplitfact}.

More interesting are the subleading power terms. We find
 \begin{align}
 A^{(1)}(1_q^+,2_{\bar{q}}^-;3_Q^+,4_{\bar{Q}}^-;5_H)_{\collin{1}{2}}&=\sqrt{x(1-x)}\bigg[\frac{\langle p4\rangle\langle r4\rangle}{\langle pr\rangle\langle 34\rangle}-\frac{[3p][3r]}{[rp][43]}\bigg]\,,\\
 A^{(1)}(1_q^+,2_{\bar{q}}^-;3_Q^-,4_{\bar{Q}}^+;5_H)_{\collin{1}{2}}&=-\sqrt{x(1-x)}\bigg[\frac{\langle p3\rangle\langle r3\rangle}{\langle pr\rangle\langle 34\rangle}-\frac{[4p][4r]}{[rp][43]}\bigg]\,, \nn
 \end{align}
and
\begin{align}
 A^{(2)}(1_q^+,2_{\bar{q}}^-;3_Q^+,4_{\bar{Q}}^-;5_H)_{\collin{1}{2}}&=\zeta \frac{xe^{i\phi}\ \langle r4\rangle^2}{2\langle pr\rangle\langle 34\rangle}+\zeta\frac{(1-x)\ e^{-i\phi}[3r]^2}{2[rp][43]}\,,
  \\
 A^{(2)}(1_q^+,2_{\bar{q}}^-;3_Q^-,4_{\bar{Q}}^+;5_H)_{\collin{1}{2}}&=-\zeta \frac{xe^{i\phi}\ \langle r3\rangle^2}{\langle pr\rangle\langle 34\rangle}-\zeta \frac{(1-x)\ e^{-i\phi}[4r]^2}{2[rp][43]}\,.
 \nn
\end{align}
These amplitudes have an interesting structure. First, note that they depend on both the $p$ and $r$ directions. These $\cO(\lambda^2)$ suppressed amplitudes have the interesting feature that they factorize as
\begin{align}\label{eq:sub_PT}
	A^{(2)}=\sum_{\hel=\pm}\text{Split}_{-\hel}(a,b;x) A_{n-1}(\ldots, r^\hel, \ldots)\,,
\end{align}
namely onto a lower point amplitude but involving the residual vector $r$. It would be interesting to understand in general the factorization structure of these amplitudes, even at tree level.  Some work in this direction at tree level has been done in \cite{Stieberger:2015kia,Nandan:2016ohb}. It seems that this depends significantly on whether or not there exists a leading power collinear limit, since \eq{sub_PT} does not hold for our earlier example. However, for our purposes, it is sufficient to be able to expand the spinor amplitudes to subleading power, whether or not a general formula can be constructed.

To compute the cross section to $\cO(\lambda^2)$, we must now consider the different interference terms, which gives a more complicated result. Noting that the color structure is identical for all helicity configurations, we simply sum over all possible configurations to get:
 
\begin{equation}
	 \sum_{\text{all configs}}|A|^2=(|A|^2)^{(0)}+(|A|^2)^{(1)}+(|A|^2)^{(2)}+\dots
\end{equation}
where the order of the various terms here is given by $(|A|^2)^{(k)}\sim \mathcal{O}(\lambda^{-2+k})$.
The leading term at $\mathcal{O}(\lambda^{-2})$ is given by:
\begin{align}
	 (|A|^2)^{(0)} &=\frac{(1-2x+2x^2)(s_{p3}^2+s_{p4}^2)-4x(1-x)s_{p3}s_{p4}}{2\zeta^2 s_{pr}s_{34}} 
   \\
   &+\frac{4x(1-x)\ \text{Re}(s_{p4}\langle p|3|r]e^{-i\phi}-s_{p3}\langle p|4|r]e^{-i\phi})^2}{\zeta^2 s_{pr}^2s_{34}^2}
 \,, \nn
\end{align} 
where $\text{Re}(z)$ denotes the real part of $z$.
The factors $\langle p|k|r]e^{-i\phi}$ turn out to appear frequently in the squared amplitudes, so it is worth getting some intuition by evaluating it explicitly.  We have that 
\begin{align}
\langle p|k|r] &= \bra{p-}\gamma_{\mu}\ket{r-} k^{\mu} \,.
\end{align} 
Using the Dirac basis for the gamma matrices, we have that 
\begin{align}
	\ket{p-}&=\frac{1}{\sqrt{2}}\begin{pmatrix}\xi\\
	-\xi
	\end{pmatrix}\ \ \ \text{ where }\xi=\begin{pmatrix}0\\ -\sqrt{p^{-}}\end{pmatrix}
  \,,\\
	|r-\rangle&=\frac{1}{\sqrt{2}}\begin{pmatrix}\eta\\
	-\eta
	\end{pmatrix}\ \ \ \text{ where }\eta=\begin{pmatrix}\sqrt{r^{+}}\\ 0\end{pmatrix}
 \nn \,.
\end{align}
Thus, it follows that 
\begin{align}
	\bra{p-}\gamma^{0}\ket{r-}&=\xi^{\dagger}\eta=0 \,,
   \\
	\bra{p-}\gamma^{i}\ket{r-}&=-\xi^{\dagger}\sigma^{i}\eta
  \,. \nn
\end{align}
This enables us to obtain the following expression
\begin{align}
	\langle p|k|r]
	&= \sqrt{p^{-}r^{+}}(k_x+ik_y) \,,
\end{align}
where $(k_x,k_y)$ are the components of the transverse momentum. 
Thus, we see that a particular simple expression follows for the following term\begin{align}
\text{Re}(\langle p|k|r]e^{-i\phi})
   &=\sqrt{p^-r^+}(k_x\cos \phi+k_y\sin\phi)
  \nn \\
&= \sqrt{s_{pr}}\ |k_{T}| (\hat{k}\cdot\hat{\phi})
\end{align}
where $|k_{T}|$ is the magnitude of the transverse momentum, and $\hat{k}$ and $\hat{\phi}$ are unit vectors in the plane transverse to $\hat{n}$. We thus gain some intuition for the appearance of the factor. Moreover, it becomes apparent that:\begin{itemize}
	\item If the expression appears linearly, it will vanish upon integration to obtain the cross section, since it is odd in $\hat{\phi}$.
	\item Secondly, it captures the effect of projecting the momentum from other sectors onto transverse components in the $n$ sector. 
\end{itemize}
We now write the $(|A|^2)^{(1)}\sim\mathcal{O}(\lambda^{-1})$ term :
\begin{equation}
\begin{split}
(|A|^2)^{(1)}=\sqrt{x(1-x)}(1-2x)
  \Bigg\{ \frac{2(s_{p3}+s_{p4})\left[|p_{3T}|(\hat{p_{3}}\cdot\hat{\phi})+|p_{4T}|(\hat{p_{4}}\cdot\hat{\phi})\right]}{ \zeta \sqrt{s_{pr}}s_{34}}\\
-\frac{4\ \left(s_{p4}|p_{3T}|(\hat{p_{3}}\cdot\hat{\phi})-s_{p3}|p_{4T}|(\hat{p_{4}}\cdot\hat{\phi})\right)(s_{p4}s_{r3}-s_{p3}s_{r4})}{\zeta s_{pr}^{3/2}s_{34}^2}
 \Bigg\}
\end{split}
\end{equation} 
which as argued in the previous part vanishes upon integration. Finally the most interesting term is the subsubleading term $(|A|^2)^{(2)}\sim\mathcal{O}(\lambda^0)$ :
\begin{align}
(|A|^2)^{(2)}
 &=1-\frac{(1-2x+2x^2)(s_{p4}s_{r3}+s_{p3}s_{r4})}{ s_{pr}s_{34}}
  \\
&+\frac{2x(1-x)(s_{p3}s_{r3}+s_{p4}s_{r4})}{s_{pr}s_{34}}+\frac{4 x(1-x)\left[|p_{3T}|(\hat{p_{3}}\cdot\hat{\phi})+|p_{4T}|(\hat{p_{4}}\cdot\hat{\phi})\right]^2}{s_{34}}
 \nn \\
&+\frac{4x(1-x) \left[s_{p4}|p_{3T}|(\hat{p_{3}}\cdot\hat{\phi})-s_{p3}|p_{4T}|(\hat{p_{4}}\cdot\hat{\phi})\right]\left[s_{r4}|p_{3T}|(\hat{p_{3}}\cdot\hat{\phi})-s_{r3}|p_{4T}|(\hat{p_{4}}\cdot\hat{\phi})\right]}{s_{pr}s_{34}^2}
 \nn \\
 &+\frac{(1-2x)^2(s_{p4}s_{r3}-s_{p3}s_{r4})^2}{s_{pr}^2s_{34}^2}
 \,. \nn
\end{align}
Since $\hat \phi$ appears quadratically here, this amplitude does not vanish when integrated over the angle $\phi$. 

Using the parametrization of this section, one can efficiently expand any amplitude expressed in terms of spinors in the two particle collinear limit. As we will show below, this is in fact sufficient to derive the leading logarithms at subleading power for event shape observables at any order in $\alpha_s$.

\section{Subleading Power Logarithms in Event Shape Observables}\label{sec:log}

Having understood how to expand spinor amplitudes in the subleading power collinear limit, we would like to apply this to the calculation of subleading power logarithms for multi-jet  observables. While our expansion techniques can be used quite generally, as an example of particular interest, we will consider the $N$-jet event shape $N$-jettiness, $\Tau_N$ \cite{Stewart:2010tn}. The $N$-jettiness observable has received significant recent attention since it can be used to formulate a subtraction scheme for performing NNLO calculations with jets in the final state, known as $N$-jettiness subtractions \cite{Boughezal:2015aha, Gaunt:2015pea}, which have been used to compute $W/Z/H/\gamma+$ jet at NNLO~\cite{Boughezal:2015dva, Boughezal:2015aha, Boughezal:2015ded, Boughezal:2016dtm,Campbell:2017dqk}, as well as inclusive photon production \cite{Campbell:2016lzl}.

\begin{figure*}[t!] 
 \centering
 \fd{8cm}{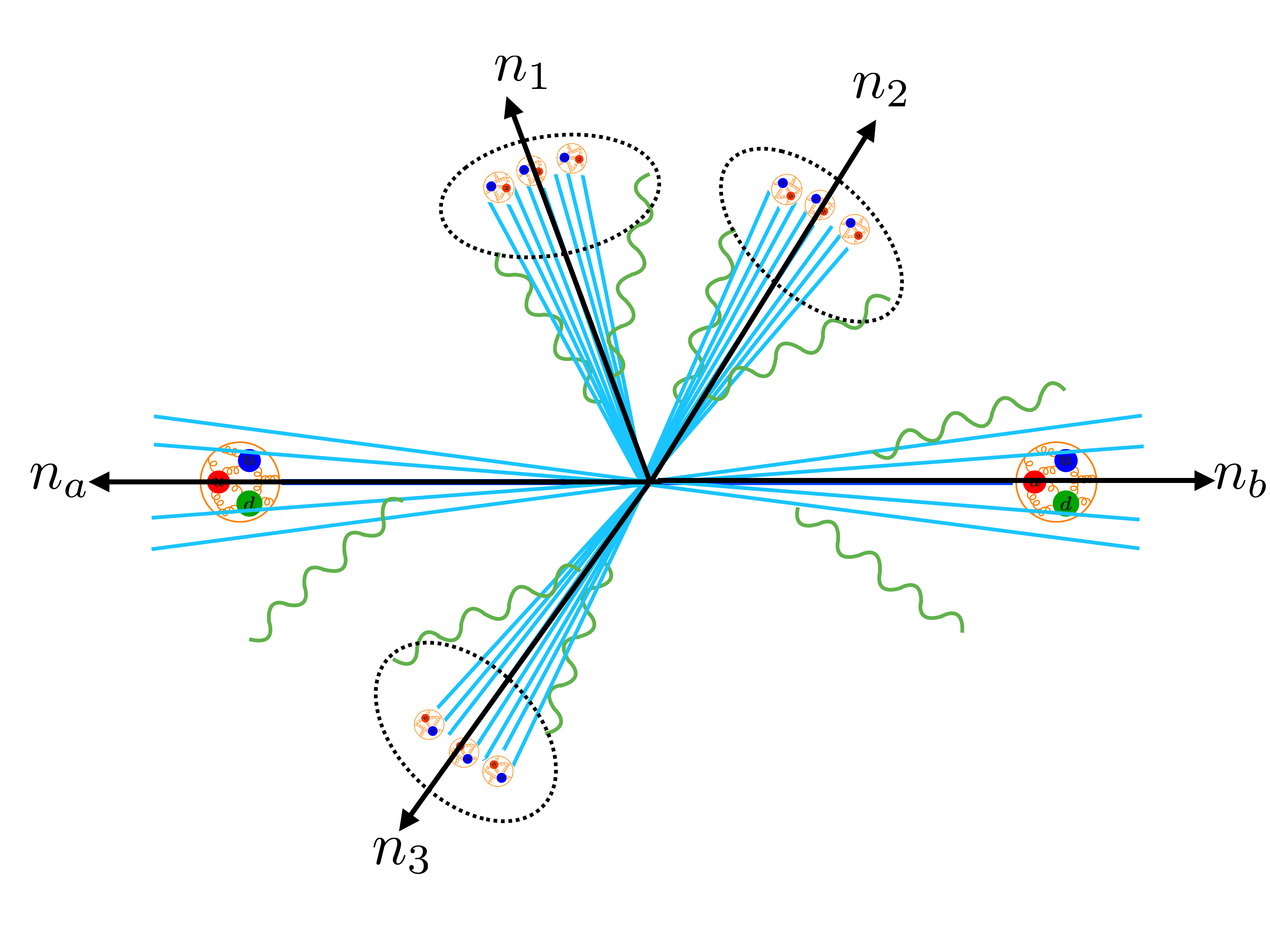}
 \caption{The $N$-jettiness observable measures the projection of radiation onto $N$ axes plus the two beam directions.}
 \label{fig:Njettiness}
\end{figure*}

The $N$-jettiness observable is defined as \cite{Stewart:2010tn}
\begin{align}\label{eq:TauN_def}
\Tau_N &= \sum_{k \in \text{event}} \min_i \Bigl\{ \frac{2 q_i\cdot p_k}{Q_i}  \Bigr\} = \sum_{j} \Tau_{N_j}
\,,\qquad  \Tau_{N_j} = \sum_{\ell \in \text{coll}_j} \frac{2 q_j\cdot p_\ell}{Q_j}
\,,\end{align}
where in the first equality the sum over $k$ runs over the total number of final state particles and the minimum runs over $i = \{a, b, 1, \ldots, N\}$.
In the remaining terms, the sum over $j$ runs over the different collinear sectors $j = \{a, b, 1, \ldots, N\}$ and the sum over $\ell \in \text{coll}_j $ runs over the number of particles in the collinear sector $j$ as determined by the minimization. This observable can therefore be viewed as projecting the radiation in the event onto $N$ axes plus the two beam directions, as shown in \Fig{fig:Njettiness}. While more general measures are possible, the above
choice $d_i(p_k) = (2q_i\cdot p_k)/Q_i$ is convenient for theoretical calculations,
because it is linear in the momenta $p_k$~\cite{Jouttenus:2011wh, Jouttenus:2013hs}.
The $q_i$ are massless reference momenta corresponding to the momenta of the hard partons
present at Born level,
\begin{equation}
q_i^\mu = E_i n_i^\mu
\,,\qquad
n_i^\mu = (1, \hat n_i)
\,,\qquad
\abs{\hat n_i} = 1
\,.\end{equation}
In particular, the reference momenta for the incoming partons are given by
\begin{equation}\label{eq:def_q}
q_{a,b}^\mu = x_{a,b} \frac{E_{\rm cm}}{2}\, n^\mu_{a,b}
\,,\qquad
n_{a,b}^\mu = (1, \pm \hat z)
\,,\end{equation}
where
\begin{align} \label{eq:beamref}
2E_a &= x_a E_{\rm cm} = n_b \cdot (q_1 + \cdots + q_N + q_L) = Q e^Y
\,, \nn \\
2E_b &= x_b E_{\rm cm} = n_a \cdot (q_1 + \cdots + q_N  + q_L) = Q e^{-Y}
\,, \nn \\
Q^2 &= x_a x_b E_{\rm cm}^2
\,, \qquad
Y = \frac{1}{2}\ln\frac{x_a}{x_b}
\,.\end{align}
Here, $q_L$ is the total momentum of any additional color-singlet particles in the Born process,
and $Q$ and $Y$ now correspond to the total invariant mass and rapidity of the Born system.
A more detailed discussion of the construction of the $q_i$ in the context of fixed-order calculations and $N$-jettiness subtractions can be found in Ref.~\cite{Gaunt:2015pea}. We will discuss this in detail below only for $1$-jettiness, which is the case of interest here.

For $\tau_N=\Tau_N/Q \ll 1$, with $Q$ a typical hard scale, one is forced into the soft and collinear limits, and one can expand the cross section in powers of $\tau_N$ as
\begin{align}
\frac{\df \sigma}{\df\tau_N}
&= \frac{\df\sigma^{(0)}}{\df\tau_N} + \frac{\df\sigma^{(2)}}{\df\tau_N}+ \frac{\df\sigma^{(4)}}{\df\tau_N} + \dotsb
\,.\end{align}
The first term in this expression, $\df\sigma^{(0)}/\df\tau_N$ contains the most singular terms, with the scaling
\begin{align}
\frac{\df\sigma^{(0)}}{\df\tau_N}
&\sim \delta(\tau_N)+ \biggl[\frac{ \ord{1} \ln^j\tau_N }{\tau_N}\biggr]_+
\,,\end{align}
with various values of $j\geq 0$.
These are referred to as the leading power terms, and a factorization formula \cite{Stewart:2009yx} describing these terms has been derived in SCET~\cite{Bauer:2000ew, Bauer:2000yr, Bauer:2001ct, Bauer:2001yt, Bauer:2002nz}. It takes the schematic form
\begin{align} \label{eq:sigma}
\frac{\df\sigma^{(0)}}{\df\tau_N} &=
\int\!\df x_a\, \df x_b\, \df \Phi_{N}(q_a \!+ q_b; q_1, \ldots)\, 
\\\nn &\quad \times
\sum_{\kappa} \tr\,\bigl[ \hH_{\kappa}(\{q_i\}) \hS_\kappa \bigr] \otimes
\Bigl[ B_{\kappa_a} B_{\kappa_b} \prod_J J_{\kappa_J} \Bigr]
\,.\end{align}
Here $B$ are beam functions, $J$ are jet functions, and $S$ is the soft function, and $\kappa$ denotes different partonic channels. The kinematic dependence on the jet directions is described by the hard function, $H$, which is the infrared finite part of the squared matrix element for the $N$-jet process. We will compare this kinematic dependence to what we find later for the power corrections.

Beyond the terms described by this factorization formula, there are terms which scale as
\begin{align}\label{eq:scaling_lam2}
\tau_N \frac{\df\sigma^{(2k)}}{\df\tau_N} &\sim \ord{\tau_N^k\, \ln^j\!\tau_N }
\,,
\end{align}
with $k\geq 1$, $j\geq 0$.
Since they are suppressed by powers of the observable, $\tau_N^k$, we refer to them as power corrections. It has been shown that the calculation of these power suppressed terms can significantly improve the performance of $N$-jettiness subtractions. This has been explicitly illustrated in the case of color singlet production in \cite{Moult:2016fqy,Boughezal:2016zws,Moult:2017jsg,Boughezal:2018mvf,Ebert:2018lzn}. However, one would like to extend this to the case of multiple jets in the final state, where they are most needed.

The calculation of the power corrections for $\Tau_N$ in the case of multiple jets is quite complicated. For applications, one would like to compute it to NNLO, namely with two additional emissions. Due to the presence of the multiple regions inherent in the $N$-jettiness definition, the multiparticle phase space becomes very complicated. Fortunately, in \cite{Moult:2016fqy} consistency relations were derived that show that the leading logarithms at subleading power can be computed to any order in $\alpha_s$ by considering virtual corrections to the subleading power two-particle collinear limits. While this was shown in the context of color singlet production, it also holds more generally, as discussed in \Sec{sec:consistency}. This implies that the parametrization of \Sec{sec:spinor} for the two particle collinear phase space is in fact sufficient to obtain the full leading log result at subleading power. This is a remarkable simplification, as it enables the calculation to be performed at any order in $\alpha_s$ as a sum over two particle collinear limits, instead of having to consider complicated multi-particle phase space integrals.

In this section we will show how to efficiently extract the leading logarithmic subleading power corrections for a process for which the helicity amplitudes are known by exploiting the consistency relations and applying the methods explained in \sec{toolbox}. In \Sec{sec:consistency} we review the consistency relations, which will allow us to derive all our results from the two-particle collinear limit. In \sec{phasespace} we set up the phase space for integrating over the two particle collinear limit, and then in \Sec{sec:log_example} we give an explicit example for $H\to q\bar q gg$. Since the goal of this paper is to illustrate the method, rather than perform a complete calculation for a particular process, we will illustrate it on simple tree level amplitudes. However, since we show that the leading logarithm can be extracted from the two particle collinear phase space, at higher orders one simply would consider the two-particle collinear limit of a more complicated amplitude.

We also note that for the complete calculation of the power corrections for the $N$-jettiness observable, one must consider not only power corrections arising from i) the expansion of the matrix element, but also power corrections arising from ii) the expansion of the phase space, ii) the flux factors, and iv) the measurement definition. These sources of power corrections, and techniques for systematically organizing their expansion have been discussed in great detail in \cite{Ebert:2018lzn}. These later three types of corrections are primarily a bookkeeping exercise, and while they do have the same importance for the final result, they are not associated with the subleading power expansion of the amplitude. In this section, we focus purely on case i), illustrating how the leading logarithms can be extracted from expanding spinor amplitudes in the two-particle collinear limits. The application to a full process of interest will be carried out in a future publication.

\subsection{Consistency Relations}\label{sec:consistency}

We begin by reviewing the consistency relations derived in \cite{Moult:2016fqy}, which will reduce the problem of computing the subleading power leading logarithms of multi-jet event shape observables to the calculation of phase space integrals over two-particle collinear limits. The results of \cite{Moult:2016fqy} were presented in the context of color singlet production, but apply more generally, since they follow from the properties of SCET amplitudes and \SCETi observables, and do not depend on the particular hard process.

We consider the fixed order calculation of the cross section in SCET. In SCET, each particle is either soft, collinear, or hard, and each graph gives a result with a homogeneous scaling in $\tau_N$, depending on the number of soft, collinear and hard particles. Explicitly, we can write the $n$-loop result for the cross section at subleading power, which we denote with the super script $(2,n)$, as
\begin{align}\label{eq:constraint_setup}
\frac{\df\sigma^{(2,n)}}{\df\tau_N}
  = & \sum_{\kappa}\sum_{i=0}^{2n-1} \frac{c_{\kappa,i}}{\epsilon^i} \left( \frac{\mu^{2n}}{Q^{2n} \tau_N^{m(\kappa)}}   \right)^\epsilon  \{  f f, f' f, f'' f, f'f' \}
\nonumber
\\
& + \dots
\,.\end{align}
Here the ellipses include  UV renormalization, and collinear PDF renormalization, which are not leading logarithmic effects, and so we will not discuss them further. In the first line, we have included a number of different PDF structures, including derivatives, which can exist in the final result. In \Eq{eq:constraint_setup} we suppress all flavor indices on the coefficients $c$, and on the PDFs, and we will continue to do this throughout this section. For example, it is implicit that the PDF structures $f'_a f_b$ and $f_a f'_b$ both occur, etc. The origin of these terms will be discussed in \Sec{sec:powlaw}. The consistency relations will hold separately for each different structure. Their arguments have been dropped, since they are not relevant for the current discussion.
In this expression $\kappa$ and $\gamma$ label the scalings obtained from the contributing particles, i.e., hard, collinear, or soft, and $m(\kappa)\geq 1$ is an integer. To be concrete, at one loop ($n=1$), we have a single particle, which can be either soft, or collinear. We therefore have
\begin{align} \label{eq:classes1}
\text{soft:} \qquad &\kappa=s\,, \qquad m(\kappa) =2\,,\nn \\
\text{collinear:} \qquad &\kappa=c\,, \qquad m(\kappa)=1 \,.
\end{align}
At two loops ($n=2$), as relevant for NNLO we have the following possibilities
\begin{align} \label{eq:classes2}
\text{hard-collinear:}\qquad  &\kappa = hc\,, \qquad m(\kappa) =1\,,\nn \\
 \text{hard-soft:} \qquad &\kappa= hs\,, \qquad m(\kappa) =2\,,\nn \\
\text{ collinear-collinear:} \qquad & \kappa=cc\,, \qquad m(\kappa) =2\,,\nn \\
 \text{collinear-soft:} \qquad & \kappa= cs\,, \qquad m(\kappa) =3\,,\nn \\
 \text{soft-soft:}  \qquad & \kappa= ss\,,\qquad m(\kappa) =4\,.
 \end{align}
 The $c_{\kappa,i}$ in \Eq{eq:constraint_setup} are coefficients of the poles in $\epsilon$ arising from the graphs with the different scalings, and differ for the different cases $\{f f, f' f, f'' f, f'f' \}$.

The main insight which allows for a dramatic simplification is that the pole terms must cancel, which places a number of relations on the values of the $c_{\kappa,i}$. In particular, at one loop, one finds
\begin{equation}\label{eq:one_loop_constraint}
c_{s,1}=-c_{c,1}
\,.\end{equation}
The leading logarithmic result at NLO for a given channel and PDF structure can then be written as
\begin{align}\label{eq:constraints_final_NLO}
\frac{\df\sigma^{(2,2)}}{\df\tau_N}
&=  \ln \tau_N ~\left(c^{(ff)}_{c,1} f f + c^{(f'\!f)}_{c,1} f' f + c^{(f''\!f)}_{c,1} f'' f+ c^{(f'\!f')}_{c,1} f'f' \right)
\,,\end{align}
implying that one need only compute either the soft contributions, or the collinear contributions.  At two loops, we have
\begin{align}\label{eq:constraints_summary}
c_{hc,3} &= \frac{c_{cs,3}}{3} = -c_{ss,3}= - \frac{1}{3} (c_{hs,3}+c_{cc,3})
\,,\end{align}
as well as a number of additional relations that were given in \cite{Moult:2016fqy}, but that are not relevant for the current discussion. 
These relations apply in each color channel, and for each combination of PDFs. For a particular contribution, we can then write the leading logarithm purely in terms of the hard-collinear coefficient
\begin{align}\label{eq:constraints_final}
\frac{\df\sigma^{(2,2)}}{\df\tau_N}
&= \ln^3 \tau_N~ \left(c^{(ff)}_{hc,3} f f + c^{(f'\!f)}_{hc,3} f' f + c^{(f''\!f)}_{hc,3} f'' f+ c^{(f'\!f')}_{hc,3} f'f' \right)
\,.\end{align}
Again, the different PDF structures in parentheses indicate that this will hold for each structure.
This implies that one need only consider a two particle collinear limit with hard virtual loops, and that no multi-particle phase space integrals need to be performed. One can simply perform the two particle phase space integral of the amplitude expanded in the two-particle collinear limit, for which we have given a convenient parametrization in terms of spinors.

\subsection{Collinear Phase Space Integral}\label{sec:phasespace}

Having shown that we can extract everything from the two particle collinear limits, in this section, we show how we can easily extract the leading logarithm at next to leading power for the matrix element corrections for the $N$-jettiness observable.
For concreteness we will consider the case of color singlet production in association with $N$-jets in $pp$ collisions, other cases like the decay of a color singlet into an arbitrary number of jets can be worked out along the same lines.

Following the notation of Ref.~\cite{Gaunt:2015pea}, to study color singlet production in association with $N$-jets in $pp$ collisions at Born level we take two incoming beams with momentum $q_a^\mu$ and $q_b^\mu$, $N$-jets with momenta $\{q_i^\mu\}_{i=1}^N$ and some non hadronic final states (the color singlet) which we take to have total momentum $q^\mu$ and we refer to the complete Born phase space $\Phi_N$ as
\be 
	\df \Phi_{N} = \frac{1}{2E_\mathrm{cm}} \int \frac{\df x_a}{x_a} \int \frac{\df x_b}{x_b} \int \df \Phi_N(q_a + q_b;q_1,\dots,q_N,q) \frac{\df q^2}{2\pi} \df \Phi_L(q)\sum_k s_k\,,
\ee 
where $ \Phi_N(q_a + q_b;q_1,\dots,q_N,q)$ is the standard $N$-body Lorentz invariant phase space, the phase space integral $ \df \Phi_L(q)$ describes the kinematics of the non-hadronic final states and $\sum_k s_k$ includes any symmetry, color and/or averaging factors, which differ for each partonic channel.
We define the \emph{born measurement} $\BornM$ to fix all kinematic variables that are not zero at leading order, hence the cross section $\df \sigma/(\df \BornM \df \Tau_N)$ which is fully differential in both the Born measurement $\BornM$ and $\Tau_N$ for such a process at LO is by definition
\be
	\frac{\df \sigma}{\df \BornM \df \Tau_N} = \int \df \Phi_{N}  |\cM|^2 \BornM(\Phi_N)  \delta(\Tau_N - \hat{\Tau}_N[\Phi_N]) = \sigma_0(\BornM) \delta(\Tau_N)\,.
\ee

Now let's consider an emission with momentum $k^\mu$ collinear to one of the collinear directions. For definiteness, let's consider an emission collinear to the $N$-th jet. We will later perform a sum over all collinear directions. The phase space for color singlet + $N$-jets and an additional emission, $\Phi_{N+1}$, can be written as a function of the born phase space for color singlet + $N$-jets, $\Phi_N$, and the two particle phase space $\Phi_2$ via
\begin{align}\label{eq:PSdecomp}
	\int \df\Phi_{N+1}(q;p_1,\dots,p_{N},k)  &= \int \df\Phi_{2}(\pstar;p_{N},k) \df\Phi_{N}(q;p_1,\dots,p_{N-1},\pstar) (2\pi)^3 \df m_0^2 \,,
\end{align}
where $m_0^2$ is the virtuality of $\pstar$ (ie. in $\df\Phi_N$ we have $\delta(\pstar^2 - m_0^2)$) and the two particle phase space\footnote{Note the $\delta^{(4)}$ that defines $\pstar^\mu = p_N^\mu +k^\mu$.} is 
\be
	 \int \df\Phi_{2}(\pstar;p_{N},k) = \int \frac{\df^4 p_N}{(2\pi)^4} \int \frac{\df^4 k}{(2\pi)^4} \delta^+(p_N^2) \delta^+(k^2) (2\pi)^4\delta^{(4)}(\pstar - p_N -k) \,.
\ee
One can then use the born measurement $\BornM$ to fix all the integrals in $\df\Phi_{N}(q;p_1,\dots,p_{N-1},\pstar) $.

If we are interested in the differential cross section in $\Tau_N$, then the phase space is constrained by the $\Tau_N$ measurement function
\be\label{eq:tau_Nmeas}
	\delta_{\Tau_N} \equiv \delta\left(\Tau_N - \hat\Tau_N[\{k,\Phi_N\}]\right) \equiv \delta\Bigl(\Tau_N - \sum_j \Tau_{N_j} \Bigr)\,,
\ee
which follows from the $\Tau_N$ definition of \eq{TauN_def}.
As explained in \cite{Stewart:2010tn}, at leading power with a single collinear emission we have 
\be 
	\Tau_{N_j}\bigr|_\text{1-emission} = t_j/Q\,, 
\ee 
where $t_j$ is the virtuality of the collinear sector $j$ where the emission lies.
In our case all $\{p_j\}_{j\neq N}$ can be taken as purely collinear such that their collinear sector has no virtuality. However, even if we choose our axis such that the $\pstar^\mu$ momentum has no perpendicular momentum, $\pstar_\perp =0$, the vector $\pstar$ still has an invariant mass. With these choices we therefore have
\be
	\Tau_{N_{j}} = 0\quad  \forall j \neq N \,, \qquad \Tau_{N_{N}} = \pstar^+\,.
\ee 
Thus the measurement takes the form
\be
	\delta_{\Tau_N} = \delta(\Tau_N - \pstar^+)\,.
\ee
Note that in general $\delta_{\Tau_N}$ can have an expansion in $\lambda$. Since we are interested only in the leading power phase space we will keep always the leading term for $\delta_{\Tau_N}$ and we will refer to it as $\delta^{(0)}_{\Tau_N}$.

The $d$-dimensional phase space for $k$ in lightcone coordinates is given by 
\begin{align}\label{eq:kPS}
	\int\frac{\df^d k}{(2\pi)^d}\delta^+(k^2) 
	&= \frac{1}{(4\pi)^2}\int \frac{\df k^+ \df k^-}{(k^+k^-)^\epsilon}  \int \frac{\df \Omega_{2-2\epsilon}}{(2\pi)^{1-2\epsilon}}\,. 
\end{align}
If the integrand is spherically symmetric we can do the solid angle integral using 
\be 
	 \int \frac{\df \Omega_{2-2\epsilon}}{(2\pi)^{1-2\epsilon}} =  \frac{(4\pi)^\epsilon}{\Gamma(1-\epsilon)} \equiv \varpi^\epsilon\,.
\ee
However, as we have seen in section \sec{spinor}, in general the amplitudes with multiple collinear directions can depend on $\phi$ at subleading power, in that case we use 
\begin{align}\label{eq:kPSphi}
	\int\frac{\df^d k}{(2\pi)^d}\delta^+(k^2) 
	&= \frac{\varpi^\epsilon}{(4\pi)^2}\int \frac{\df k^+ \df k^-}{(k^+k^-)^\epsilon}  \int_0^{2\pi} \frac{\df \phi}{2\pi}\,.
\end{align}
We call $Q$ the energy (or large component) of the jet momentum $\pstar$ and $x$ the fraction of it that the emission takes away. In this way we change variable 
\be\label{eq:changeofvarx}
	k^- \to x\, Q\,\,\,,\qquad\int_0^Q \frac{\df k^-}{(k^-)^\epsilon} = Q^{1-\epsilon}\int_0^1 \frac{\df x}{x^\epsilon}
\ee
and one can show
\footnote{ 
	From \eq{PSdecomp} use the $\df m_0^2$ integral and the $\delta(\pstar^2 - m_0^2)$ which is part of $\df\Phi_N$, to solve the $\delta(p_N^2)$ after using momentum conservation:
	\be 
		\int \df m_0^2 \underbrace{\delta(m_0^2 - 2\pstar \cdot k)}_{\text{from }\delta^+(p_N^2)} \underbrace{\delta(\pstar^2 - m_0^2)}_{\text{part of $\Phi_N$}}\underbrace{\delta(\Tau_N - \pstar^+)}_{\delta_{\Tau_N}} =\underbrace{\delta(\pstar^2 - \Tau_N \pstar^-)}_{\text{put it back into }\Phi_N} \delta(k^+ - \Tau_N(1-x))
		\,.
	\ee
}
that the $N$-jettiness measurement fixes the $k^+$ component via
\be\label{eq:deltaT}
	\delta_{\Tau_N}^{(0)} = \delta\left( k^+ - (1-x)\Tau_N\right)\,.
\ee
Combining \eqs{changeofvarx}{deltaT} we get
\be
	\int \frac{\df k^+ \df k^-}{(k^+k^-)^\epsilon} \delta^{(0)}_{\Tau_N} = Q(\Tau_N Q)^{-\epsilon} \int_0^1 \frac{\df x}{x^\epsilon(1-x)^{\epsilon}}\,.
\ee
Therefore, the leading power phase space for $N$-jets + one collinear emission inside the $i$-th jet reads%
\begin{align}\label{eq:collinearphasespacesingle}
	\int \df\Phi_{N+1}\delta_{\Tau_{N_i}}&= \int \df \Phi^{(0)}_N\, \left(\frac{\varpi}{\Tau_{N} Q}\right)^{\epsilon}\int_0^1 \frac{\df x}{x^\epsilon(1-x)^{\epsilon}} \int_0^{2\pi} \frac{\df \phi}{(2\pi)}  \frac{Q}{(4\pi)^{2}}  + \cO(\Tau_{N}) \nn\\
	&\equiv \int \df \Phi^{(0)}_N\, \df \Phi^{(0)}_{2,i}(\Tau_{N}) + \cO(\Tau_{N})\,,
\end{align}
where we defined
\be
	\df\Phi^{(0)}_{2,i}(\Tau_{N}) \equiv \left(\frac{\varpi}{\Tau_{N} Q}\right)^{\epsilon}\int_0^1 \frac{\df x}{x^\epsilon(1-x)^{\epsilon}} \int_0^{2\pi} \frac{\df \phi}{(2\pi)}  \frac{Q}{(4\pi)^{2}}\,,
\ee
as the two particle phase space resulting from one emission inside the $i$-th jet constrained by the $\Tau_N$ measurement.
Note that in general the $\Tau_N$ measurement \eq{tau_Nmeas} gets contributions from the radiation $k^\mu$ being collinear to \emph{any} collinear direction in the event. Since we are considering only one emission on top of the Born configuration at a time we are always able to isolate the contribution to the $\Tau_N$ measurement to one collinear sector, hence the leading power phase space for $N$-jets + one collinear emission inside any jet reads%
\be\label{eq:collinearphasespacemultiple}
	\int \df\Phi_{N+1}\delta\left(\Tau_N - \sum_j \min_i \Bigl\{ \frac{2 q_i\cdot p_j}{Q_i} \Bigr\}\right)= \int \df\Phi_{N}^{(0)} \sum_{i=1}^{N} \Phi_{2,i}(\Tau_{N})\,.
\ee

A special case to consider separately is when the emission is collinear to one of the beams which contains the incoming particles. 
In that case the parton distribution functions enter the collinear phase space. 
For concreteness let's take $k^\mu\parallel q_a^\mu$. The steps follow closely those already done above so we won't repeat them. 

The main difference is that, since%
\footnote{
	Note that this relation is true only at LP, which is enough for the LP phase space calculation we are considering here. If one were to consider the power corrections coming from the phase space, then in general $q_a^\mu = x_a \Ecm \biggl(\frac{1}{z_a}+ \underbrace{\Delta_a^{(2)}}_{\sim \Tau_N} \biggr) \frac{n^\mu}{2}$. See Section 3 of Ref. \cite{Ebert:2018lzn} for a detailed discussion on this.} 
$q_a^\mu = \frac{x_a}{z_a} \Ecm \frac{n^\mu}{2}$, we have
\be
	 \delta_{\Tau_{N_a}}^{(0)} = \delta(\Tau_{N} - e^{-Y} k^+)\,,\qquad k^- = x_a \frac{(1-z_a)}{z_a}\,,
\ee
and one makes the choice of using a different change of variable for $k^-$ namely $k^- = x_a \Ecm \frac{(1-z_a)}{z_a}$.
In this way the PDF related to the beam direction to which $k^\mu$ is collinear enters the collinear integral, and we have
\be
	\df\Phi_{2,a}(\Tau_{N_a}) f_a(x_a)= \left(\frac{\varpi}{\Tau_{N} Q}\right)^{\epsilon} \int_{x_a}^1 \frac{\df z_a}{z_a } f_i\left(\frac{x_a}{z_a}\right) \frac{z_a^\epsilon}{(1-z_a)^{\epsilon}} \int_0^{2\pi} \frac{\df \phi}{(2\pi)}  \frac{Q}{(4\pi)^{2}}\,,
\ee
where the $f_a(x_a)$ factor is needed to keep the same normalization as in \eq{collinearphasespacesingle}.

We now want to combine \eq{collinearphasespacemultiple} with the power correction to the matrix element squared in order to get the subleading component of the fully differential cross section due to the matrix element correction. 
In order to do so, let us define the matrix element squared expansion as two particles go collinear in the $q_\colllabel$ as
\be
	|\cM|^2\, =\, \underbrace{\Msquared_\colllabel^{(0)}}_{\sim \lambda^{-2}}\, +\, \underbrace{\Msquared_\colllabel^{(2)}}_{\sim \lambda^{0}} \,+\, \cO(\lambda)\,,
\ee 
and we have assumed that $\Msquared_\colllabel^{(1)}$ vanishes upon integration over $\phi$.
We also note that $\Msquared^{(i)}$ is only a function of the born variables contained in $\BornM$, $\Tau_N$, and $x,\phi$
\be
	\Msquared^{(i)}=\Msquared^{(i)}(\BornM,\Tau_N,x,\phi)\,.
\ee 
In the following, we are going to leave understood the explicit dependence on $\BornM$ in $\Msquared^{(i)}$. 

Using the result for the leading power phase space, the subleading component of the fully differential cross section due to the matrix element correction for an emission inside a jet reads
\begin{align}\label{eq:XSCollinearJet}
	\frac{\df\sigma_{\Msquared,\text{jet}}^{(2)}}{\df \BornM \df \Tau_N} &= \sum_{k k^\prime=\{q,g\}}\frac{f_k(x_a)f_{k^\prime}\left(x_b\right)}{2\Ecm^4 x_a x_b} \sum_{\colllabel=1}^N \int \df \Phi_{2,\colllabel}\Msquared_{\colllabel}^{(2)}(x,\phi)\\
	&= \sum_{k k^\prime=\{q,g\}}\frac{f_k(x_a)f_{k^\prime}\left(x_b\right)}{2\Ecm^4 x_a x_b} \sum_{\colllabel=1}^N \left(\frac{\mu^2}{\Tau_N Q}\right)^{\epsilon}\int_0^1 \frac{\df x}{x^\epsilon(1-x)^{\epsilon}} \int_0^{2\pi} \frac{\df \phi}{(2\pi)} Q\left(\frac{\alpha_s}{4\pi}\right) \Msquared_{\colllabel}^{(2)}(x,\phi)\,,\nn
\end{align}
where we extracted for convenience the coupling and its $\overline{\text{MS}}$ scale $\mu$ from the matrix element squared since we will use helicity amplitudes which are typically given with the coupling understood.
We conclude by considering the case where the radiation $k^\mu$ is collinear to one of the beams, in this case we have
\begin{align}\label{eq:XSCollinearBeam}
	\frac{\df\sigma_{\Msquared,\text{beam}}^{(2)}}{\df \BornM \df \Tau_N} &= \sum_{k k^\prime=\{q,g\}} \frac{f_{k^\prime}\left(x_b\right)}{2\Ecm^4 x_a x_b} \left(\frac{\mu^2}{\Tau_N Q}\right)^{\epsilon} \int_{x_a}^1 \frac{\df z_a}{z_a } f_k \left(\frac{x_a}{z_a}\right) \frac{z_a^\epsilon}{(1-z_a)^{\epsilon}}
   \\
 &\times \int_0^{2\pi} \frac{\df \phi}{(2\pi)} Q\left(\frac{\alpha_s}{4\pi}\right) \Msquared^{(2)}_a(z_a,\phi)
  + (a \leftrightarrow b)\,, \nn 
\end{align}
where $x_{a,b}$ are Born variables fixed by $\df \BornM$ and $\Msquared_{a(b)}^{(2)}$ is the one emission matrix element squared when the radiation is collinear to the incoming $n$($\bn$) direction (and we remind the reader that the subscript $(2)$ indicates that this is the subleading power term in the collinear expansion which is suppressed by $\cO(\lambda^2)$ w.r.t. the leading term).
Together \eqs{XSCollinearJet}{XSCollinearBeam} give the necessary expressions to obtain the leading logarithms.

\subsection{Example of Extracting Subleading Power Logarithms}\label{sec:log_example}

In this section we give two examples of extracting the logarithms of the event shape. 
This is meant to serve two purposes. 
First, it will illustrate how simple it is to extract logarithms of the event shape from the two-particle collinear limit. 
Second, although we will not compute the complete result for the power corrections for $N$-jettiness for a particular process, the results will illustrate the general structure appearing in such results, which is already quite interesting.

\subsubsection{Full Matrix Element Corrections for $gq \to Hq$}\label{sec:example_Tau0}
Let us start by analyzing the case of Higgs production in gluon fusion at next to leading power. In this case the power correction to the fully differential cross section has been calculated at LL in Ref. \cite{Moult:2017jsg} both at NLO and NNLO.\footnote{The full $\cO(\alpha_s)$ correction has later been computed in Ref.~\cite{Ebert:2018lzn}.  Earlier results for the inclusive cross section in the hadronic $\tau$ definition can also be found in Refs. \cite{Boughezal:2016zws,Boughezal:2018mvf} at LL and NLL respectively.} Reproducing the contribution to this result coming from the matrix element corrections will give us the occasion to illustrate the techniques presented in this paper in a known example and to cross check the result. Note that we will be following the notation of \cite{Ebert:2018lzn} where the separation of the phase space contributions and the matrix element corrections are given explicitly.

The master formula for the matrix element corrections to the fully differential cross section of color singlet production in the collinear limit is given by\footnote{Taking for simplicity the leptonic definition of $\Tau$, $\rho = e^Y$} \cite{Ebert:2018lzn}
\begin{align} \label{eq:sigma_NLO_NLP_coll_FULL}
 \frac{\df\sigma_{n, \Msquared^{(2)}}^{(2)}}{\df Q^2 \df Y \df\Tau} &
 = \int_{x_a}^1 \frac{\df z_a}{z_a} \,
   \frac{f_a\left(x_a/z_a\right)\, f_b(x_b)}{2 x_a x_b \Ecm^4}
   \frac{ z_a^\eps}{(1-z_a)^\eps}\frac{\bigl(Q \Tau\bigr)^{-\eps} Q(4\pi)^{\eps}}{\Gamma(1-\eps) (4\pi)^2 }\Msquared^{(2)}(Q, Y, z_a)
\,,\end{align}
where the Born measurement has been chosen to be the color singlet invariant mass $Q$ and rapidity $Y$.
Note that \eq{sigma_NLO_NLP_coll_FULL} correctly matches \eq{XSCollinearBeam} up to our slightly different conventions here for the inclusion of coupling, $\overline{\text{MS}}$ scale and factors in $\Msquared^{(2)}$.
We now want to apply the techniques of \sec{spinor} to compute $\Msquared^{(2)}(Q, Y, z_a)$ and extract the leading logarithmic term by taking the soft limit of the amplitude and plug it in \eq{sigma_NLO_NLP_coll_FULL}. For conciseness we limit ourselves to the $gq \to Hq$ channel.\\

The relevant amplitudes are \cite{Schmidt:1997wr} 
\begin{align}
	A(1^+,2_q^+,3^-_{\bar{q}};4_H) &= -\frac{1}{\sqrt{2}}\frac{[12]^2}{[23]}\,, \nn\\
	A(1^-,2_q^+,3^-_{\bar{q}};4_H) &= -\frac{1}{\sqrt{2}}\frac{\langle 13\rangle^2}{\langle 23\rangle}\,.
\end{align}
Now, we implement our expansion, and square to obtain 
\begin{align}\label{eq:qgMsq}
	\Msquared^{(0)}_{gq,n}=0\,,\quad	\Msquared^{(2)}_{gq,n}=2C_F\frac{\Msquared^{\text{LO}}}{Q^4}\frac{(1-x)^2\ s_{p2}}{x}+\mathcal{O}(\lambda)\,,
\end{align}
where $\Msquared^{\text{LO}}$ is the LO amplitude squared for $gg\to H$ and $x$ is the momentum fraction of the quark as it becomes collinear with the gluon. Note that if any entity is incoming, then in our formalism, we need to replace all components $p^-\rightarrow -p^-$, and take $|p\rangle\rightarrow i\ |p\rangle$. Doing this ensures that we match the condition of being positive $p^0$. Therefore, since in this example we are taking the $p_1^\mu$ and $p_2^\mu$ momenta to be incoming, we need to implement the following changes:\begin{align}
	x&=\frac{p_3^-}{-p_1^-+p_3^-}\,,\quad
	p^-=-p_1^-+p_3^-\,,
\end{align}
and the Mandelstam invariant appearing in \eq{qgMsq} takes the form
\be
		s_{p2}=p^-p_2^+=-(p_1^--p_3^-)p_2^+ = -Q^2(1+\cO(\lambda^2))\,.
\ee
To compare our result with the notation of Ref. \cite{Ebert:2018lzn}, where the cross section is expressed in terms of the splitting variable $z_a$, such that $k^- = Qe^Y\frac{1-z_a}{z_a}$, we need
\begin{align}
	x=1-\frac{1}{z}\,.
\end{align}
Using $ x=1-\frac{1}{z}$ and $s_{p2}=-Q^2$, \eqref{eq:qgMsq} reads
\be
	\Msquared^{(0)}=0\,,\qquad\Msquared^{(2)}=\frac{2C_F \Msquared^{\text{LO}}}{Q^2 z(1-z)}\,,
\ee
which matches Eq.(5.22) of \Ref{Ebert:2018lzn} up to the difference in the normalization convention and $\cO(\epsilon)$ terms. Given that also the phase space \eq{sigma_NLO_NLP_coll_FULL} matches \eq{XSCollinearBeam}, this is sufficient to reproduce the leading log for this channel at subleading power.
\subsubsection{A simple $H+1$ jet example: $H\to \bar q q \bar Q Q$}\label{sec:example_HqqQQ}
We now move to the more involved case of multiple collinear directions and consider the simple example of $H\to \bar q q \bar Q Q$. When a quark $q$ and an antiquark of different flavor $\bar{Q}$ become collinear, the squared subleading power amplitudes are given by \Eq{eq:hqqQQsq}
\begin{align}
|A(1_q^+,2_{\bar{q}}^-;3_Q^+,4_{\bar{Q}}^-;5_H)|_{\collin{1}{4}}^2&=\frac{\left[(1-x)\ s_{p2}+x\ s_{p3}\right]^2}{4s_{p2}s_{p3}\ x(1-x)}\,,\\
|A(1_q^+,2_{\bar{q}}^-;3_Q^-,4_{\bar{Q}}^+;5_H)|_{\collin{1}{4}}^2&=\frac{s_{23}^2}{4 s_{p2}s_{p3}\ x(1-x)}\,.
\end{align} 
Integrating over the two particle collinear phase space, we have
\begin{align}
\left(\frac{\mu^2}{\Tau_N Q}\right)^{\epsilon}\!\!\int_0^1 \frac{\df x}{x^\epsilon(1-x)^{\epsilon}}|A(1_q^+,2_{\bar{q}}^-;3_Q^+,4_{\bar{Q}}^-;5_H)|_\collin{1}{4}^2  &=\left( \frac{s_{p2}}{4s_{p3}}  + \frac{s_{p3}}{4s_{p2}} \right)\left[  -\frac{1}{\epsilon}+\ln \frac{Q \Tau_N}{\mu^2}  + \text{finite} \right]  ,\nn \\
\left(\frac{\mu^2}{\Tau_N Q}\right)^{\epsilon} \!\!\int_0^1 \frac{\df x}{x^\epsilon(1-x)^{\epsilon}}  |A(1_q^+,2_{\bar{q}}^-;3_Q^-,4_{\bar{Q}}^+;5_H)|_\collin{1}{4}^2  &=  \frac{s_{23}^2}{2 s_{p2}s_{p3}} \left[  -\frac{1}{\epsilon}+\ln \frac{Q \Tau_N}{\mu^2} + \text{finite}  \right].
\end{align}
Here we see that we are able to easily extract the $1/\epsilon$ divergence, or correspondingly the logarithm. Already from this simple example, we can see an interesting, but expected result, namely that at subleading power the kinematic dependence on the jet directions will no longer be that of the $N$-jet process, as was true in the leading power factorization in \Eq{eq:sigma}. This will be interesting to study numerically for a complete process, and is also important phenomenologically as it controls the rapidity dependence of the power corrections.

We have therefore shown that we can efficiently extract subleading power logarithms in event shape observables from the subleading power collinear limits. Even in the case of complicated multi-leg amplitudes, the ability to extract the entire logarithm from the two-particle collinear limit means that there is only a single angular integral, which even if it cannot be done analytically, is finite, and therefore can be done by numerically evaluating the spinors. This should be achievable in a fairly automated way, enabling power corrections to be computed for multi-jet event shapes.

\section{Power Law Divergences in Subleading Power Matrix Elements}\label{sec:powlaw}

In this section we wish to elaborate on an interesting physical effect observed in the expansion of multi-point amplitudes at subleading power, namely the appearance of power law singularities. As an example, we can consider the $H\to  ggq\bar q$ amplitude at subleading power. For simplicity, we can focus on the color stripped amplitude for a particular helicity configuration\cite{Kauffman:1996ix}
\begin{equation}
A(1^+,2^-,3_q^+,4_{\bar{q}}^-;5_H)=\frac{\langle 24\rangle^3}{\langle 12\rangle\langle 14\rangle\langle 34\rangle}-\frac{[13]^3}{[12][23][34]}\,,
\end{equation}
and we can consider the behavior in the collinear limit when the gluon 1 and the quark 3 become collinear. This collinear limit has no leading power term, and using the expansion in \eq{deffinal} we find that the subleading power term in the expansion is given by
\begin{equation}
A(1^+,2^-,3_q^+,4_{\bar{q}}^-;5_H)_\collin{1}{3}=0\times \cO(\lambda^{-1}) + \frac{\langle 24\rangle^3}{x\sqrt{1-x}\ \langle p2\rangle\langle p4\rangle^2}+\mathcal{O}(\lambda)\,,
\end{equation}
where $x$ parametrizes the energy fraction of the gluon.
Squaring the amplitude, we find
 \begin{equation}
|A_{ggq\bar{q}H}|_\collin{1}{3}^2= \frac{s_{24}^3}{x^2(1-x)s_{p2}s_{p4}}+\mathcal{O}(\lambda)\,,
\end{equation}
which is $\cO(\lambda^2)$ suppressed with respect to the leading power and exhibits a $1/x^2$ divergence as the gluon becomes soft. This is distinct from the behavior of the leading power splitting functions, which go like $1/x$. These divergences are of course regulated by dimensional regularization in the phase space integral. However, their treatment requires the use of distributional identities which are less familiar than those required to treat the more standard $1/x$ divergences. For this reason, here we discuss briefly how these divergences are treated for a SCET$_{\rm I}$ type observable (like N-jettiness), and the manner that they appear for different processes and a wider class of observables.

In the case that the two collinear particles are both in the final state, one can simply integrate over all values of $x$ in the standard fashion. The power law divergences is then  regulated by dimensional regularization. More interesting, is when the subleading power collinear limit arises from an initial state splitting. In this case one has an integral against the PDFs, and one must expand as a distribution to extract the divergence, as is familiar at leading power. At subleading power we encounter a wider class of distributions, beyond the common $\delta$-function and $+$-functions.  The impact of these power law divergences on PDFs has been discussed in detail in Ref.~\cite{Ebert:2018gsn}, so here we only provide a brief review that suffices for our discussion here.

Consider the integral
\begin{align} \label{eq:toy_dist2}
 I_m = \int_0^1 \df x\,\frac{\tilde g(x)}{(1-x)^{1+m+\epsilon}} \,,
\end{align}
where $m\ge 0$ is an integer.
Here we have put the divergence at $x\to 1$, as is standard in the parameterization of initial state splittings, and $\tilde g(x)$ contains, for example, the PDFs, and other functions that are regular as $x\to 1$.
For $m=0$, the divergence can be extracted using the familiar distributional identity
\begin{align} \label{eq:plus_dist_1}
 \frac{1}{(1-x)^{1+\epsilon}} = - \frac{\delta(1-x)}{\epsilon} + \cL_0(1-x) + \cO(\epsilon)
\,.\end{align}
Here $\cL_0(1-x) = [ 1/(1-x) ]_+^1$ is the standard plus distribution, which satisfies
\begin{align}
 \int_x^1 \df x\, \tilde g(x) \cL_0(1-x) = \int_x^1 \df x \, \frac{\tilde g(x) - \tilde g(1)}{1-x} + \tilde g(1) \underbrace{\int_x^1 \df x \, \cL_0(1-x)}_{=\ln(1-x)}
 \,,\quad x \in [0,1]
\,.\end{align}
This standard plus distribution is sufficient for the treatment of divergences encountered at leading power.

At subleading power, where $m>0$, this must be generalized to multiple plus distributions. For the particular case of $m=1$, we have
\begin{align} \label{eq:double_plus}
  \frac{1}{(1-x)^{2+\epsilon}} &
 = \frac{\delta'(1-x)}{\epsilon} - \delta(1-x) + \cL_0^{++}(1-x) + \cO(\epsilon)
\,.\end{align}
Here we encounter the double plus function, whose action on a function is given by
\begin{align}
 \int_x^1 \df x \, \tilde g(x) \cL_0^{++}(1-x) &
 = \int_x^1 \df x \frac{\tilde g(x) - \tilde g(1) - \tilde g'(1)(x-1)}{1-x}
 \nn\\*&\quad
 + \tilde g(1) \underbrace{\int_x^1 \df x\, \cL_0^{++}(1-x)}_{-x/(1-x)}
 +\, \tilde g'(1) \underbrace{\int_x^1 \df x\, (x-1)  \cL_0^{++}(1-x)}_{-\ln(1-x)}
\,.\end{align}
Most importantly, we find that the divergence is associated with a $\delta'(1-x)$, instead of the standard $\delta(1-x)$ that is familiar at leading power. When integrated against the PDFs, this will give derivatives of the PDFs, an interesting physical effect which occurs at subleading powers. 

Derivatives of PDFs also appeared in the calculation of subleading power corrections to $\Tau_0$, which were computed in  \cite{Moult:2016fqy,Boughezal:2016zws,Moult:2017jsg,Boughezal:2018mvf,Ebert:2018lzn}. In this case the derivatives of the PDFs 
have a very simple origin, arising from residual momentum routed into the PDFs. Namely, one finds power corrections arising from the expansion of the PDFs
\begin{equation}
f_i\biggl[\xi\Bigl(1 + \frac{k}{Q}\Bigr)\biggr] = f_i(\xi) + \frac{k}{Q}\, \xi f_i'(\xi) + \dotsb
\,,\end{equation}
where $k\sim \cO (\lambda^2)$. This is quite different than the case found here, where the $1/x^2$ singularity arises from the structure of the amplitude itself rather than from expanding kinematics.

\begin{table}[t!]
	\centering
		\begin{tabular}{|c|c|c|c|c|}
			\hline
			&\multicolumn{2}{c}{\bf \small Color singlet}\vline & \multicolumn{2}{c}{\bf \small Color singlet + 1 jet}\vline\\
			\cline{2-5}
			&\SCETi {\scriptsize ~($\tau$,$\Tau_N$)}&\SCETii {\scriptsize ~($p_T$)}&\SCETi {\scriptsize ~($\tau$,$\Tau_N$)}&\SCETii {\scriptsize ~($p_T$)}\\
			\hline
			&&&&\\[-3mm]
			$(|A|^2)^{(0)}(x)$& {\Large $\frac{1}{x}$} & 1& {\Large $\frac{1}{x}$}  &1 \\[3mm]
			$(|A|^2)^{(2)}(x)$& {\Large $\frac{1}{x}$} & {\Large $\frac{1}{x}$} & \red{\Large ${\frac{1}{x^{2}}}$} &? \\[3mm]
			$\Phi^{(0)}(x)$& 1 & {\Large $\frac{1}{x}$} & \red{1} &{\Large $\frac{1}{x}$} \\[3mm]
			$\Phi^{(2)}(x)$& 1 & {\Large $\frac{1}{x^2}$} & ? & ? \\[3mm]
			\hline
		\end{tabular}
	\caption{Behavior of the most singular terms for matrix element squared contributions $(|A|^2)^{(k)}$ and phase space $\Phi^{(k)}$,  at both leading power $k=0$ and subleading power $k=2$, for a single emission when the energy fraction of the emission $x\to 0$. The contribution to the cross section involves products of these terms: $\df \sigma^{(i+j)} \sim A^{(i)} \times \Phi^{(j)}$, see \eq{dfsigmadec}.
	In red we highlight the terms analyzed in this paper. We take the explicit results for $A^{(2)}$ and $\Phi^{(2)}$ for fully differential color singlet production in \SCETi from \cite{Ebert:2018lzn}, while those for \SCETii are taken from \cite{Ebert:2018gsn}. Entries with `` ? " have not yet appeared in the literature.}
   \label{tab:singularities}
\end{table}

More generally one can ask the question whether power law divergences are a general feature of subleading power calculations in SCET. To answer this question one can analyze the behavior of the cross section at next to leading power both for \SCETi (where examples include jet masses, thrust, beam thrust) and \SCETii measurements (where examples include observables with a small transverse momentum $q_T\ll Q$). The NLP corrections to the cross section can be schematically described at one emission as 
\be\label{eq:dfsigmadec}
	\df \sigma^{(2)} \sim\int_0^1 \frac{\df x}{x}\left[ A^{(0)}(x) \Phi^{(2)}(x) + A^{(1)}(x) \Phi^{(1)}(x) + A^{(2)}(x) \Phi^{(0)}(x)\right]\,,
\ee
where $A^{(0)}$ is the leading power matrix element squared, $\Phi^{(0)}$ is the leading power phase space and $A^{(i)}$ or $\Phi^{(i)}$ indicate the $i$-th order in the $\lambda$ expansion of $A$ or $\Phi$ respectively.

In \Tab{tab:singularities} we summarize the behavior of the most singular terms both for the phase space and the matrix element squared in different contexts, varying the process and the type of observable.
In the case of a process with only 2 back-to-back directions (like color-singlet production with no additional jets) and an \SCETi measurement, Refs.~\cite{Moult:2016fqy,Boughezal:2016zws,Moult:2017jsg,Boughezal:2018mvf,Ebert:2018lzn} found no power law divergences both at LL and NLL. 

However, for the same color single process but with an \SCETii ($p_T$) measurement, Ref.~\cite{Ebert:2018gsn} found power law divergences at subleading power. This includes observables such as the $p_T$ spectrum of a Higgs or Drell-Yan case.  This is indicated by the entries in the second column of \Tab{tab:singularities}. 
In this case the $1/x^2$ singularity arises from the subleading power phase space, $\Phi^{(2)}$, or from the product $(|A|^2)^{(2)}\Phi^{(0)}$ which each scale like $1/x$ individually. 

The example considered here is given in the third column of \Tab{tab:singularities}. Here the result is different, since the power law divergence occurs directly in the expansion of the matrix element itself, $(|A|^2)^{(2)}$. This is a feature that, as far as we know, has never been encountered before in the literature. Again these divergences are regulated by the use of dimensional regularization.  
The appearance of power law singularities directly from the expansion of the amplitude for an observable with an additional jet is quite interesting, and we expect to be a generic property of the $N$-jet case. 

For the case of $H\to gg$, the subleading power logarithms exponentiate to all orders with the double logarithms governed by the cusp anomalous dimension \cite{Moult:2018jjd}. Loosely speaking, this follows from the fact that the expansion in the soft and collinear limits inherited its properties from the leading power expansion. In the more general $N$-jet case, due to the presence of the power law singularities, we expect that this will no longer be the case, and it will be of significant interest to understand the all orders structure of the subleading power logarithms.

\section{Conclusions}\label{sec:conclusions}

In this paper we have shown how we can use spinor-helicity amplitudes to efficiently study subleading power collinear limits and extract logarithms of infrared observables in high multiplicity final states.  Our approach uses consistency relations derived from effective field theory to show that the leading logarithm in the subleading power expansion of the event shape observable can be derived from the subleading power expansion of the two-particle collinear limit, for which we gave an efficient parametrization in terms of spinor variables. This approach significantly simplifies the analysis, as it avoids the need to consider complicated multi-particle phase space integrals.

In our extension to higher point amplitudes, we have noticed some interesting features of the power expansion. In general, for higher point amplitudes, we find that power law divergences are present in subleading power squared amplitudes themselves. These lead to observable effects, in particular, the appearance of derivatives of the PDFs. Derivatives of the PDFs have appeared in the calculation of power corrections in other contexts, and we discussed and contrasted the different mechanisms in \sec{powlaw}.

We believe that there are a number of immediate phenomenological applications of our techniques. In particular, our techniques can be applied to compute the power corrections for $N$-jettiness subtractions for $H/W/Z+$ jet. These are the processes for which the power corrections are most needed from a phenomenological perspective, but have so far been too cumbersome to compute. Using our consistency relations, the power corrections up to NNLO can be computed using the one-loop $H/W/Z+4$ parton amplitudes, all of which are known analytically in terms of spinors \cite{Bern:1996ka,Bern:1997sc,Berger:2006sh,Badger:2007si,Badger:2009vh,Dixon:2009uk, Badger:2009hw}. It would also be interesting to understand in more detail the subleading power collinear behavior of general amplitudes. We intend to pursue these directions in future work.

\begin{acknowledgments}
We thank Franz Herzog for providing us several simplified squared amplitudes which we used as cross checks of our techniques, and Markus Ebert for discussion of distributional identities related to multiple plus distributions. This work was supported in part by the Office of Nuclear Physics of the U.S.
Department of Energy under Contract No. DE-SC0011090, by the Office of High Energy Physics of the U.S. Department of Energy under Contract No. DE-AC02-05CH11231, and by the Simons Foundation Investigator Grant No. 327942. This work was also partially supported by the UROP Endowment Fund, MIT.
\end{acknowledgments}

\bibliography{bibliography}{}
\bibliographystyle{jhep}

\end{document}